%% file: REUltrafast150113.tex
\documentclass[pra,superscriptaddress,reprint]{revtex4-1}
\usepackage{amsmath}
\usepackage{hyperref}
\usepackage{wasysym}
\usepackage[squaren]{SIunits}

\usepackage{graphicx,color}

\definecolor{myorange}{rgb}{1,.5,0} 
\definecolor{mygreen}{rgb}{0,0.5,0} 
\definecolor{myblue}{rgb}{0,0,0.75} 
\definecolor{mymagenta}{cmyk}{0,1,0,0.12}

\input{NewCommands.tex}

\newcommand{\erf}{{\rm erf}}

\newcommand{\di}{d^{(\rm i)}}
\newcommand{\ds}{d^{(\rm s)}}
\newcommand{\dl}{d^{(\rm l)}}

\renewcommand{\u}{{p}}

\newcommand{\uone}{{p^{({\rm s})}}}
\newcommand{\utwo}{{p^{({\rm l})}}}
\newcommand{\udet}{{p^{({\rm i})}}}
\newcommand{\phic}{\phi^{(\rm c)}}
\newcommand{\phiq}{\phi^{(\rm q)}}
\newcommand{\sigq}{\sigma_q}
\newcommand{\phidet}{\phi_{\rm det}}

\newcommand{\vis}{{\cal V}}
\newcommand{\vel}{V^{(\rm el)}}
\newcommand{\vprev}{V^{(\rm prev)}}
\newcommand{\error}{\vprev}

\newcommand{\Pred}{{\cal P}}
\newcommand{\meanPred}{\overline{\rm Pred}}
\renewcommand{\meanPred}{\overline{\Pred}}
\newcommand{\meanPredwc}{\meanPred^{(\worst)} }

\begin{document}


\title{Strong experimental guarantees in ultrafast quantum random number generation} 



\author{Morgan W.~Mitchell}
\email[]{morgan.mitchell@icfo.es}
\affiliation{ICFO-Institut de Ciencies Fotoniques, Av. Carl Friedrich Gauss, 3, 08860 Castelldefels, Barcelona, Spain.}
\affiliation{ICREA-Instituci\'{o} Catalana de Recerca i Estudis Avan\c{c}ats, 08015 Barcelona, Spain}

\author{Carlos Abellan}
\affiliation{ICFO-Institut de Ciencies Fotoniques, Av. Carl Friedrich Gauss, 3, 08860 Castelldefels, Barcelona, Spain.}

\author{Waldimar Amaya}
\affiliation{ICFO-Institut de Ciencies Fotoniques, Av. Carl Friedrich Gauss, 3, 08860 Castelldefels, Barcelona, Spain.}

\date{\today}

\begin{abstract}
{
We describe a  methodology and standard of proof for experimental claims of quantum random number generation (QRNG), analogous to well-established methods from precision measurement.  For appropriately constructed physical implementations, lower bounds on the quantum contribution to the average min-entropy can be derived from measurements on the QRNG output.  Given these bounds, randomness extractors  allow generation of nearly perfect ``$\epsilon$-random'' bit streams.  An analysis of experimental uncertainties then gives experimentally derived confidence levels on the $\epsilon$ randomness of these sequences.  We demonstrate the methodology by application to phase-diffusion QRNG,  driven by spontaneous emission as a trusted randomness source.  All other factors, including classical phase noise, amplitude fluctuations, digitization errors and correlations due to finite detection bandwidth, are treated with paranoid caution, i.e., assuming the worst possible behaviors consistent with observations.   A data-constrained numerical optimization  of the distribution of untrusted parameters  is used to lower bound the average min-entropy.  Under this paranoid analysis, the QRNG remains efficient, generating at least 2.3 quantum random bits per symbol with 8-bit digitization and at least 0.83 quantum random bits per symbol with binary digitization, at a confidence level of  0.99993.  The result demonstrates ultra-fast QRNG with strong experimental guarantees.}

\end{abstract}


\maketitle 

%


\newcommand{\AME}{\overline{H}_{\rm \infty}}
\newcommand{\AMEd}{\tilde{H}_{\rm \infty}[d]}
\renewcommand{\AMEd}{\widetilde{H}_{\rm \infty}}
\newcommand{\AMEdet}{\overline{H}_{\rm \infty}[d]}
\renewcommand{\AMEdet}{\overline{H}_{\rm \infty}}

\section{Introduction}

Quantum random number generation extracts randomness from quantum mechanical processes and measurements.  Processes used have included radioactive decay \cite{YoshizawaJJSCS1999}, path-splitting of single photons \cite{Jennewein2000}, photon number path entanglement \cite{Kwon2009}, amplified spontaneous emission \cite{Williams2010}, measurement of the phase noise of a laser \cite{Guo2010, Jofre2011, Xu2012, AbellanOE2014},  photon arrival time \cite{WahlAPL2011}, vacuum-seeded bistable processes \cite{MarandiOE2012} and stimulated Raman scattering \cite{BustardOE2013}.  Quantum random number generators are attractive because their randomness can be linked to well-tested principles of quantum mechanics, e.g.  the uncertainty principle \cite{VallonePRA2014}, which guarantees a minimum amount of randomness in some physical quantities.

{
Physics plays an essential role in QRNG, not only at the generation stage, but also when making claims of randomness.   While it is common to test generated data against statistical test suites \cite{Rukhin2010}, these tests can only identify nonrandomness, i.e., patterns in the output.  For fundamental reasons, statistical tests cannot confirm randomness of finite sequences \cite{FrauchigerARX2013}.  In contrast, physical models can support a randomness claim, as we describe in this work.  

Trust plays a central role in contemporary discussions of QRNG, as it does in quantum cryptography.  Cryptography employs {trust models} that define what parts of a communication system are assumed to be understood, in contrast to those that could be under the control of an adversary.  A strategy that trusts fewer parts of the system places a lower burden on verification.  In an extreme of paranoia, ``device-independent'' (DI) strategies distrust even the measurement devices employed by the communicating parties \cite{AcinPRL2007,AbbottMSC2014,LiARX2014,BancalNJP2014,WangPRA2014}.  
The DI approach aims to provide security against hardware-based attacks \cite{GerhardtNComms2011}, and some progress toward DI QRNG has been demonstrated \cite{PironioN2010}.  

It is important to note that DI techniques aim to guarantee considerably more than randomness.  They use loophole-free Bell inequality violations \cite{AcinPRL2007}, or other evidence for nonlocality \cite{LiARX2014,BancalNJP2014}, in conjunction with monogamy relations and the no-signaling principle to guarantee that no other actor could be in possession of a copy of the generated random numbers.  This guarantee has obvious security value and explains much of the interest in DI quantum key distribution and DI QRNG.  In practice, however, loophole-free Bell inequality violations are experimentally difficult, and the demonstrated rates are very low.  A heroic experiment that still left open the timing loophole produced 42 random bits in 1 month \cite{PironioN2010}, 15 orders of magnitude slower than other techniques \cite{AbellanOE2014,YuanAPL2014}.  For the foreseeable future, practical use of QRNGs will require verification.  Moreover, many randomness applications, e.g. Monte Carlo simulations, have no reason to protect themselves against information leakage and obtain no benefit from the additional security of the DI approach.

Nearly all experimental claims of QRNG to date implicitly or explicitly assume nonadversarial devices, with varying degrees of trust in their sources \cite{Jennewein2000,Kwon2009,KanterNP2010,Guo2010,WayneOE2010,Jofre2011,WahlAPL2011,SymulAPL2011,LiOL2012,MarandiOE2012,Argyris2012,Xu2012,WangOE2013,BustardOE2013,YuanAPL2014,VallonePRA2014}.  To take the best-known example, splitting a single photon on an ideal 50:50 beam-splitter gives a random direction to the photon, and this direction can be measured to give one perfectly random bit.  DI-grade paranoia is not practical in this scenario; if the beam-splitter transmission were under the control of an adversary, she could determine every outcome.  It is thus necessary to verify the performance of the device.  Unfortunately, most QRNG claims, indeed all that we are aware of, leave  important gaps in the verification.  In the beam-splitter example, a variety of classical effects could steer the outcome: correlations in the photon source, inefficiency in the detectors, light entering the unused port of the interferometer, sensitivity of the beam-splitter to polarization, frequency, beam position, beam direction, or any other variable that might fluctuate in the light source, to name a few.  Some of these effects, e.g. variable detector efficiency \cite{StefanovJMO2000,WeiOL2009}, have been accounted for, while others  have not.  For continuous-variable (CV)  QRNGs, a category that includes the fastest devices, the accounting for noise and detection bandwidth has to date been unrealistically optimistic.  For example, it is often assumed that digitization noise is independent of the quantum noise being digitized \cite{GabrielNPhot2010,Xu2012} or that detection systems introduce no correlations \cite{SymulAPL2011,SanguinettiPRX2014}.  As we show in Sec. \ref{sec:digitization}, these assumptions are unwarranted in real systems.  Concerning analysis, only a few experimental works \cite{Xu2012,BustardOE2013,AbellanOE2014} quantify their performance using measures compatible with modern randomness extraction (see Sec. \ref{sec:RandomnessQuantification1}).




We propose a standard of proof for quality assurance in QRNG, between the paralyzing ``trust-nothing'' paranoia of the DI approach and the risky insouciance of most QRNG demonstrations to date.    
We refer to this as {\em metrology-grade paranoia}.  The name notes the similarity of the verification required for characterization of a QRNG  and the verification required to make a precision measurement.  Both practices assume that the system is fundamentally understandable, but take a conservative and rigorous approach to calibration and experimental imperfections, i.e., to systematic errors.  A modern precision measurement, e.g., of the transition frequency in an atomic clock, will take into account a large variety of possible systematic errors and give a quantitative estimation of their effect on the measurement result \cite{maddaloniBook2013,BloomN2014}.  Both approaches burden the experimenter with understanding and quantifying all relevant aspects of their system.  The success of similar approaches in precision measurement reassures us that this burden is not unbearable. 

We apply our approach to phase-diffusion QRNG \cite{Jofre2011, Xu2012}, the fastest reported QRNG approach \cite{AbellanOE2014,YuanAPL2014}.  We show that the statistics of the measured output provide lower bounds on the amount of quantum randomness contained in the data stream, allowing the generation of $\epsilon$-random sequences and the assignation of confidence levels to the purity of the randomness.  We find that the claims for pulsed phase diffusion survive metrology-grade paranoia, and thus it is possible to have simultaneously a very high bit rate and strong randomness assurance in a practical system.  

}

\newcommand{\Gbps}{Grps~}
\newcommand{\conv}{{*}}
\newcommand{\cA}{A}
\newcommand{\cB}{B}
\newcommand{\cC}{C}
\newcommand{\cFF}{F}
\newcommand{\trans}{{\cal T}}
\newcommand{\refl}{r}
\newcommand{\CDF}{{F}}
\newcommand{\CDFu}{{F_{\circ} }}
\newcommand{\digital}{D}
\newcommand{\minent}{H_{\rm min}}
\newcommand{\xvec}{{\bf{x}}}
\newcommand{\yvec}{{\bf{y}}}
\newcommand{\pvec}{{\bf{p}}}
\newcommand{\supup}{^{({\rm u})}}
\newcommand{\suplo}{^{({\rm l})}}
\newcommand{\supupideal}{^{({\rm u,ideal})}}
\newcommand{\suploideal}{^{({\rm l,ideal})}}
\newcommand{\supupmax}{^{({\rm u,max})}}
\newcommand{\suplomax}{^{({\rm l,max})}}
\newcommand{\supupmin}{^{({\rm u,min})}}
\newcommand{\suplomin}{^{({\rm l,min})}}
\renewcommand{\uplus}{u_+}
\newcommand{\umin}{u_-}
\newcommand{\Freq}{F_{\rm rel}}

\begin{figure}[t]
\includegraphics[width=0.43 \textwidth]{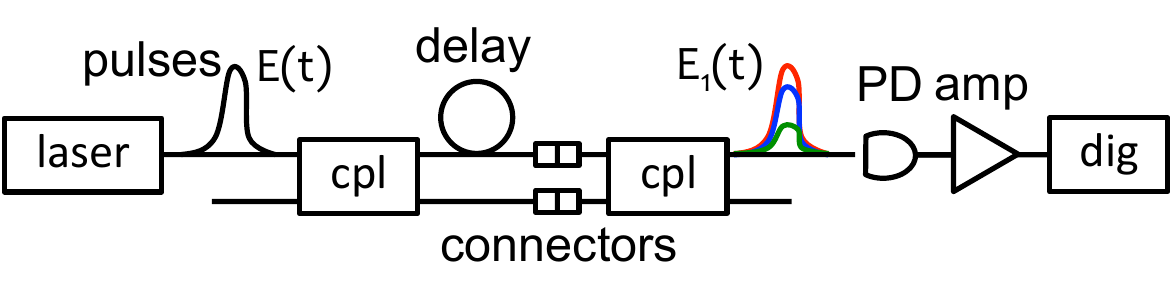} \\
\includegraphics[width=0.43 \textwidth]{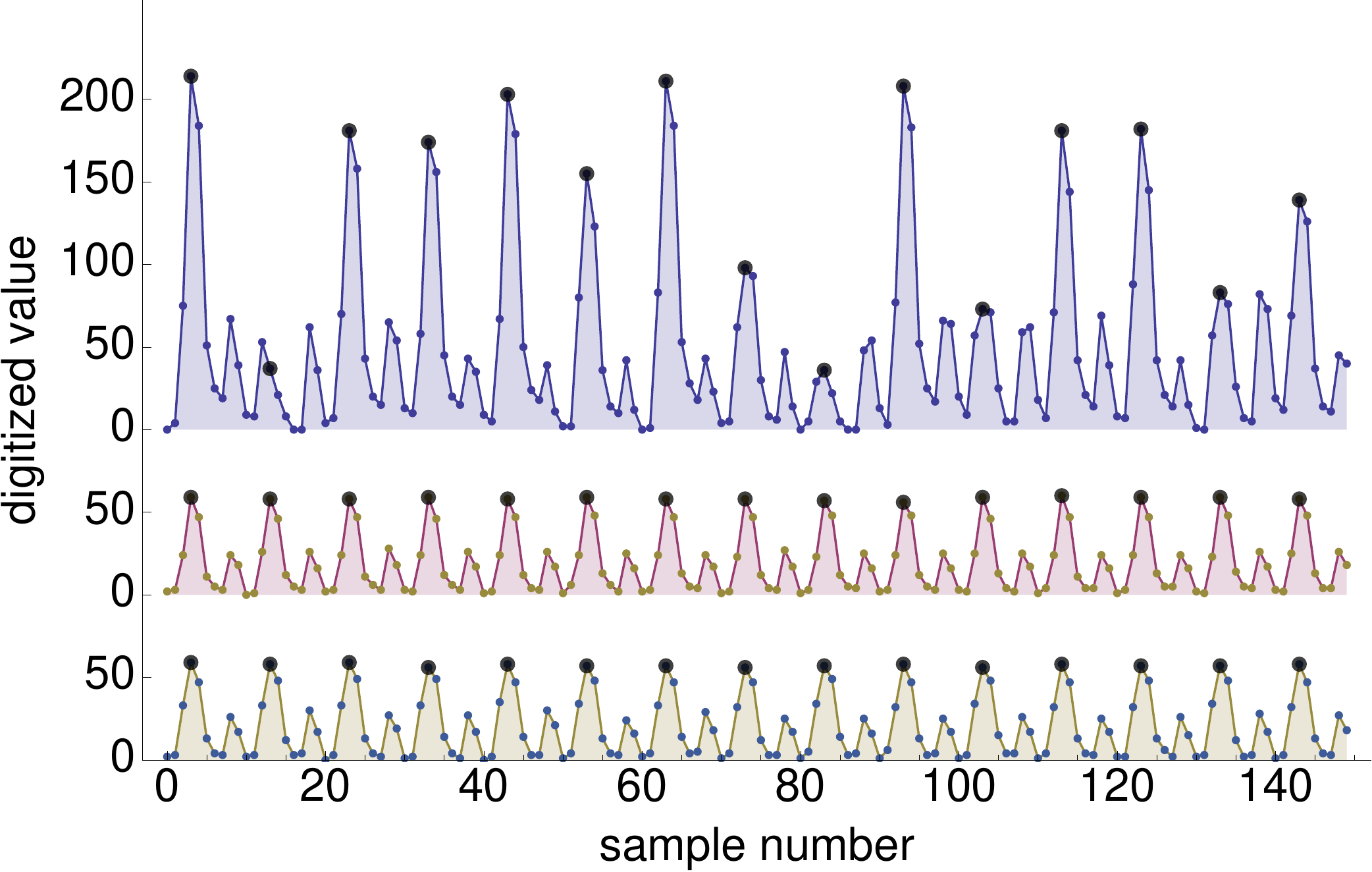} \\ 
\includegraphics[width=0.43 \textwidth]{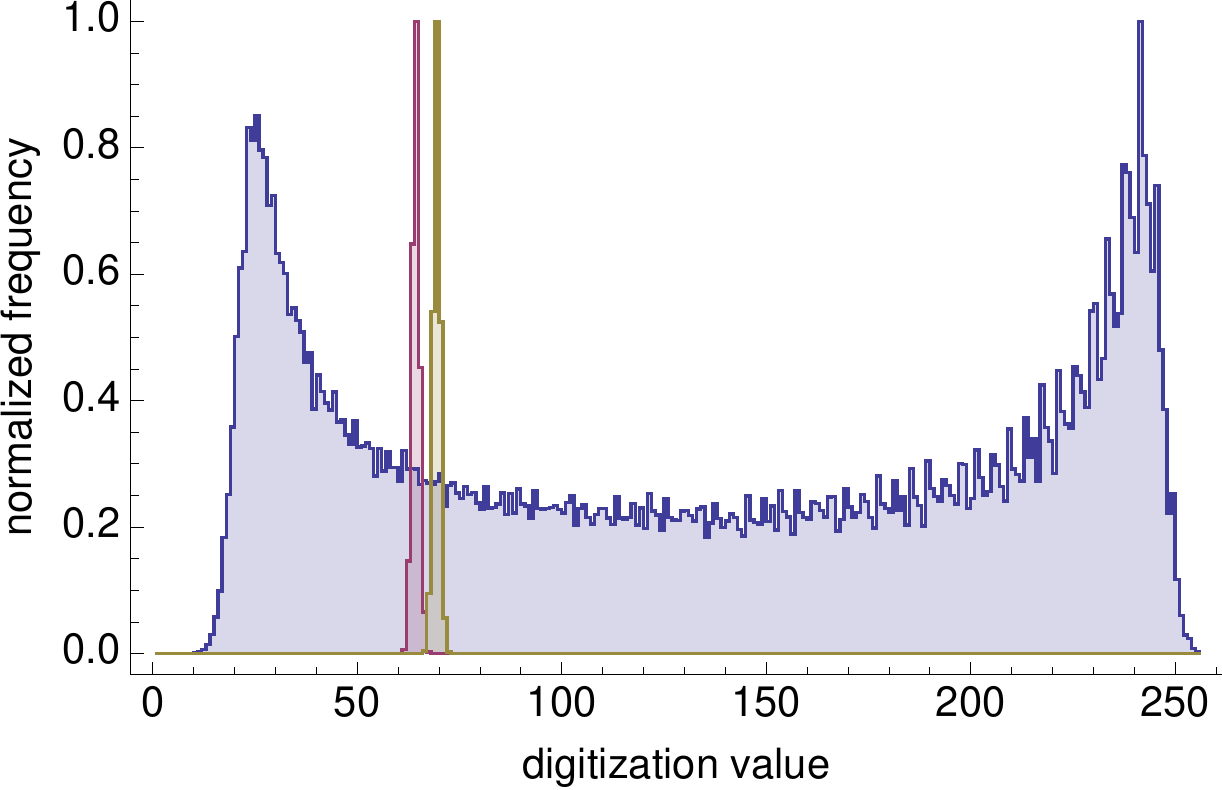}
\caption{ (Color online) (Top)  Schematic of phase-diffusion QRNG.  A single-mode diode laser is strongly current modulated to produce a train of phase-randomized output pulses with field strengths $E(t)$.  Interference of subsequent pulses is performed with a Mach-Zehnder interferometer, consisting of single-mode $2 \times 2$ couplers (cpl) and a relative delay equal to the pulse-repetition period $\tau$.  A photodiode (PD) converts the 
output pulse powers into electrical current, which is amplified (amp) and converted to digital values with a digitizer (dig).  Either arm of the MZI can be broken to measure the pulse amplitude in the other arm.  (Middle)  Time domain recording of a short digitized sequence of $\udet$, the interferometer output with interference (top, blue), and $\uone$ (middle, red), and $\utwo$ (bottom, beige), the outputs of the interferometer with only the 
short or long path open, respectively.   Data have been shifted to have equal baselines.  (Bottom) Histograms (scaled for equal height) for $\udet$ (wide, blue), $\uone$ (left narrow, red), and $\utwo$ (right narrow, beige).  
The wide $\udet$ distribution arises from interference and resembles the arcsine distribution that describes $\cos \phi$ when $\phi$ is uniformly distributed. 
 }
\label{fig:schematic}
\label{fig:realtime}
\label{fig:histos}
\end{figure}

\begin{figure}[tr]
\includegraphics[width=0.48 \textwidth]{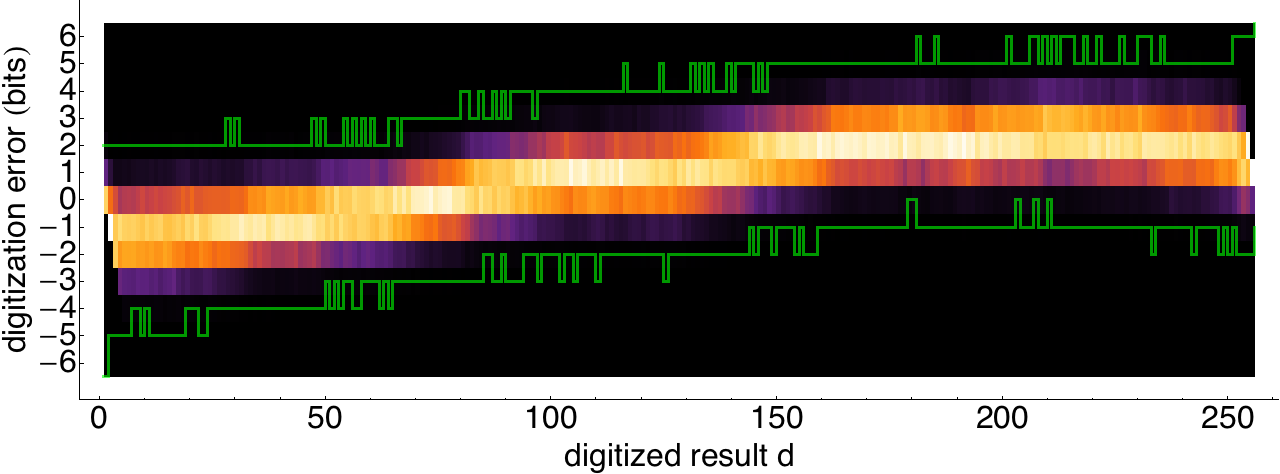} 
\caption{(Color online) Measured digitization error frequencies and error limits.  Color indicates relative frequency from zero (black) to maximum (white).  It is interesting to note the presence of both a large-scale nonlinearity in the conversion (the general trend) and small-scale regularities (e.g. the period-two patterns clearly visible between 50 and 60). Green traces above and below indicate the largest and smallest errors observed, respectively.   Approximately $2^{14}$ samples per digitization value were used to obtain the frequencies, so the confidence that a new event will fall within the limits is $\approx 1- 2^{-14}$.  }
\label{fig:digilimits}
\end{figure}

\newcommand{\cov}{{\rm cov}}
\newcommand{\BW}{{\rm BW}}

\section{Randomness quantification}
\label{sec:RandomnessQuantification1}

A perfect physical device is not required for near-perfect randomness generation.  Algorithms known as randomness extractors (REs) \cite{Nisan1999, FrauchigerARX2013} convert partly random data into nearly perfect ``$\epsilon$-random'' bit strings by a hashing process \footnote{In the tradition of {\em qubit} from ``quantum bit'' and {\em ebit} from ``entanglement bit'' or ``entangled bit,'' we humbly propose the term {\em rabit}, from ``random bit,'' as the binary unit of disinformation \cite{ViswanathMC1998}}.  If $d$ is a random symbol with probability distribution $P(d)$, then ${\cal P} \equiv \max_{d} P(d)$ is the predictability, and $H_{\rm \infty} \equiv - \log_2 {\cal P}$ is the min-entropy.  Information-theoretically provable REs \cite{TrevisanJACM2001,FrauchigerARX2013} can produce $\epsilon$-random output bit strings with a length given by their input min-entropy.  

Real devices do not operate under constant conditions, and it is necessary to accommodate the possibility that a QRNG is at some moments producing higher-quality randomness than at other moments.  We can describe this situation saying the symbol $d$ has a probability distribution $P(d|\xvec)$, where $\xvec$ describes the condition of the source when $d$ is produced.  Although $\xvec$ may vary, it is not a source of true randomness.  It describes parameters not trusted to be random; for example, the $\xvec$ variation may be deterministic but unknown to us.  We consider the randomness quantification from the perspective of someone, perhaps an adversary, who knows $\xvec$.  Because $\xvec$ includes all of the untrusted variables, and because the trusted variables are independent, subsequent $d$ are independent, in the sense that the probability $P(\{d\}| \{\xvec \})$ of generating a string of output symbols $\{ d \} \equiv (d_1, \ldots , d_N)$ under conditions $\{ \xvec \} \equiv (\xvec_1, \ldots , \xvec_N)$ is given by the product
$P(\{d\}| \{\xvec \}) =  \Pi_i P(d_i|\xvec_i) $.
The conditional {min-entropy} of $\{ d \}$ is then
\begin{equation}
H_\infty( \{d\} | \{ \xvec \} ) \equiv - \log_2 \min_{\{d\} } P(\{d\})  = \sum_i H_{\rm \infty}(d_i|\xvec_i)
\end{equation}
where $H_{\rm \infty}(d|\xvec) \equiv - \log_2 \min_d P(d|\xvec)$ is the conditional min-entropy of a single symbol generated with conditions $\xvec$.  Note that $H_\infty( \{d\} | \{ \xvec \})$ does not depend on the order of the elements of $\{\xvec \}$, so that a knowledge of the relative frequencies $\Freq(\xvec)$ with which the conditions $\xvec$ appear in $\{ \xvec \}$ is sufficient to compute the mean min-entropy per symbol,
\begin{equation}
\label{eq:AMEdet}
\AMEdet =  \int d\xvec\,  \Freq(\xvec) H_\infty(d|\xvec).
\end{equation}
As we shall see, a measured string $\{d\}$, combined with a model of how $\xvec$ and trusted randomness interact in the source to produce $d$, constrain $\Freq(\xvec)$, and thus provide a bound on $\AMEdet$ {\em for that string}.  In this way, randomness guarantees, with no prior assumptions about $\{ \xvec \}$, can be generated, at the cost of analyzing each raw string $\{d\}$.  

If we allow ourselves to assume that the conditions $\{ \xvec \}$ are independent random variables \footnote{This assumption is not as strong as it may appear.  Consider a scenario in which $P(d)$ depends on an environmental variable, e.g. temperature, that is correlated over some finite time scale.  We can choose to define $d$ as the sequence of raw outcomes produced during a time longer than the correlations, and $\xvec$ as the corresponding sequence of environmental variables.  This restores the independence condition, at the cost of a larger alphabet of symbols.  }, it suffices to characterize $P(\xvec)$, the distribution of $\xvec$, rather than $\Freq(\xvec)$, the relative frequencies that actually occur.   REs adapted to this probabilistic situation  \cite{dodis2004fuzzy,DodisSIAMJoC2008}  give $\epsilon$-random output with length limited by the {\em average min-entropy}, defined as
\begin{equation}
\label{eq:AMEdef}
\AMEd \equiv - \log_2 \int d\xvec \, P(\xvec)  \max_d P(d|\xvec).
\end{equation}
Note the difference relative to Eq. (\ref{eq:AMEdet}); here the logarithm is outside of the average.  This reduces the entropy, so that for $P(\xvec) = \Freq(\xvec)$, $\AMEd \le \AMEdet $. 
As with $\Freq(\xvec)$ and $\AMEdet$, $P(\xvec)$ and $\AMEd$ can be bounded using knowledge of a measured string $\{ d\}$, but this calculation only needs to be performed once, and can be performed with a very long string $\{d\}$, to precisely estimate $P(\xvec)$.  

In what follows, we work with $P(\xvec)$ and $\AMEd$, the more conservative of the two entropy measures, although the same methods can be applied to $\Freq(\xvec)$ and $\AMEdet$.

\section{methodology}
In principle, the prescription for metrology-grade paranoia is simple.   First, describe the process by which a quantum random variable, in our case $\phi_q$, the laser phase diffusion due to spontaneous emission, and other experimental variables $\xvec$ combine to produces measurement results $d$.     Second, use the distribution of $\phi_q$, known from first principles or from modeling, to calculate $P(d|\xvec)$, the distribution of symbols $d$, conditioned on $\xvec$.  Third, find $\underline{\AMEd}$, the lowest value of $\AMEd$ that is consistent with what is known about $\xvec$, i.e., with experimental or theoretical constraints on $P(\xvec)$, the distribution of $\xvec$.  

Knowing $\underline{\AMEd}$, a RE can then be used to produce an $\epsilon$-random bit string, with length $\approx  N  \underline{\AMEd}$, where $N$ is the number of symbols in the raw data string. 
Confidence in the randomness of this bit string derives from the confidence in $P(\xvec)$.  For example, if statistical and systematic uncertainties give 99\% confidence  that the process produced at least $\underline{\AMEd}$ average min-entropy, then the extracted bit string is $\epsilon$-random with at least that same confidence level. 

The consistency condition is an invitation to paranoia.  For example, it has sometimes been assumed in QRNG work that digitization errors are independent of the quantum signal being digitized, and simply add entropy to the raw data, an entropy that is not of quantum origin and must be accounted for in order to not overestimate the quantum entropy, but is otherwise harmless.  But is this really the case ?  How can one be sure that the noise added by the digitizer is {\em independent} of the signal ?  Unless one possesses specific knowledge about this characteristic of the digitizer in question, one must admit that our knowledge is {\em consistent} with less favorable scenarios \cite{ SymulAPL2011}.  For example, the digitizer might organize its errors to bias the results toward one subset of possible symbols, reducing the entropy and in effect consuming some of the quantum randomness present.   A paranoid analysis must assume this is indeed happening, and in the way that reduces  $\AMEd$ as much as possible. 

To show that this methodology can be used in practice, we perform this analysis on a phase-diffusion QRNG, of the same design as \cite{AbellanOE2014}.

\section{Model}

We start with the model shown in Fig. \ref{fig:schematic}~{(top)}, corresponding to  \cite{Jofre2011,AbellanOE2014}.    A single-mode diode laser is driven with a strongly modulated injection current with period $\tau$.  For all data shown in this work, $\tau = 5$~ns.   The optical output of the laser, described by the field $E(t)$, is fed to the input of an unbalanced Mach-Zehnder interfereometer (MZI), with short and long delays $\tau_s$ and $\tau_l = \tau_s + \tau$, respectively. The field exiting the MZI is
\bea
E_{1}(t) &=& \trans_{s} E(t-\tau_{s}) + \trans_{l} E(t-\tau_{l}),
\eea
where $\trans_{s}$ and $\trans_{l}$ are the  transmission coefficients, including both couplers, for the short and long paths, respectively.  A photodiode converts the incident power, $\udet(t) = |E_1(t)|^2$, into a current, which is amplified and digitized at times $t_i = i \tau$, $i=1,2,\ldots$ with the time origin chosen near the peak of the pulse.  Due to strong phase-diffusion between times $t_i$ and $t_{i+1}$, the detected signal shows a strong variation that is not present in the input pulses.  This is illustrated in Fig. \ref{fig:realtime}~{(middle)}, which shows digitized signals, both from the complete MZI with interference, and from the MZI with either arm interrupted.  Histograms of the resulting interference and single-path signals are  shown in Fig. \ref{fig:histos}~{(bottom)}.

The phase between pulses contains a quantum contribution $\phiq$ as well as a classical contribution $\phic$, due to relative phase of the interferometer arms, as well as classical fluctuations in laser parameters such as injection current.  {As described in the Appendix, 
quantum theory of laser dynamics  \cite{Agrawal1990,ScullyZubairy1997} predicts that $\phiq$ is independently distributed from one pulse to the next, with} a Gaussian probability density function (PDF) $P(\phiq)$ of rms width $\sigq$.  We keep $\sigq$ as a parameter, in order to study its effect on randomness generation. 
Writing the total phase $\phic(t) + \phiq(t) = \arg E(t - \tau_s) - \arg E(t-\tau_l)$ and suppressing time dependencies for clarity, the optical signal, i.e., the instantaneous power,  is 
\bea
\label{eq:interference}
\udet &\equiv& \uone + \utwo + 2 \vis \sqrt{\uone \utwo} \cos (\phic + \phiq),
\eea
where  $\uone(t) \equiv | \trans_{s} E(t-\tau_{s})|^2$, $\utwo(t) \equiv | \trans_{l} E(t-\tau_{l})|^2$, and $\vis(t)$ is the interference visibility.  We assume the photodetection and amplification process is linear and stationary, so the electrical signal arriving to the digitizer is 
\bea
\label{eq:V}
V(t) &=& \int_{-\infty}^{t} dt' \, G(t-t') \udet(t') + \vel(t)
\eea
where $G$ is the impulse response of the detector-amplifier-digitizer system and $\vel$ is the summed electronic noise from all sources.  Finally, the digitizer converts $V$ to a digital value $d$. 
Digitization is a highly nonlinear process, and requires special care, as we now describe.

\newcommand{\ac}{{\rm ac}}
\newcommand{\cc}{{\rm cc}}
\newcommand{\HinfQ}{H_{\rm \infty}^{({\rm Q})}}

\section{digitization}
\label{sec:digitization}
Fig. \ref{fig:histos}~{(bottom)} illustrates a feature of digitization.  This process adds classical noise, e.g. from the amplification, and moreover employs a highly nonlinear electronic operation to convert a continuum of inputs $\udet$ into a finite set of outputs $d$.  Although it may be tempting to assume that errors in this process are independent of $\udet$ (as is typically the case for amplifier noise), this is clearly untrue for digitization noise.  For example, a digitizer will normally have a measurable preference for even versus odd outputs \cite{KesterBook2005}, something that would not occur if errors were independent of the input.  In Fig. \ref{fig:histos}, an oscillation in the histogram frequencies with period 4 is clearly visible, with an amplitude that is modulated with a period of 16. These errors have an rms width of $0.8$ {codes, i.e., increments of the digitizer output,} when averaged over all $d$, and are clearly not independent of $\udet$. 

We experimentally bound the size of digitization errors as follows.  We use an electronic function generator (Tabor WW1281A) followed by a low-pass filter to produce a quasistatic voltage (a 1-kHz triangle wave) and digitize this signal with our fast 8-bit digitizer (Acqiris U1084A) and simultaneously with a 14-bit oscilloscope (Agilent infiniium 86100C with an electronic module Agilent 86112A)  for reference.  
Fig. \ref{fig:digilimits} shows the distribution of digitization errors, i.e., of the deviation of the digitized value from the ideal value, based on $\approx 2^{14}$ samples per digitization value.  This allows us to identify limits  $V_d^{(\min)}$ and $V_d^{(\max)}$, the minimum and maximum voltages, respectively, that were observed to produce a given digitization value $d$.  Below, to compute a lower bound on $\HinfQ$ in the presence of digitization errors, we  assume that digitization results outside of these limits are so improbable as to have a negligible effect on $\HinfQ$.  We note that electronic noise during the characterization measurements, e.g., in the voltage source or in the reference oscilloscope, can only broaden these bounds, making them conservative.

\newcommand{\acdel}{\Delta}

\begin{figure}[tr]
\includegraphics[width=0.48 \textwidth]{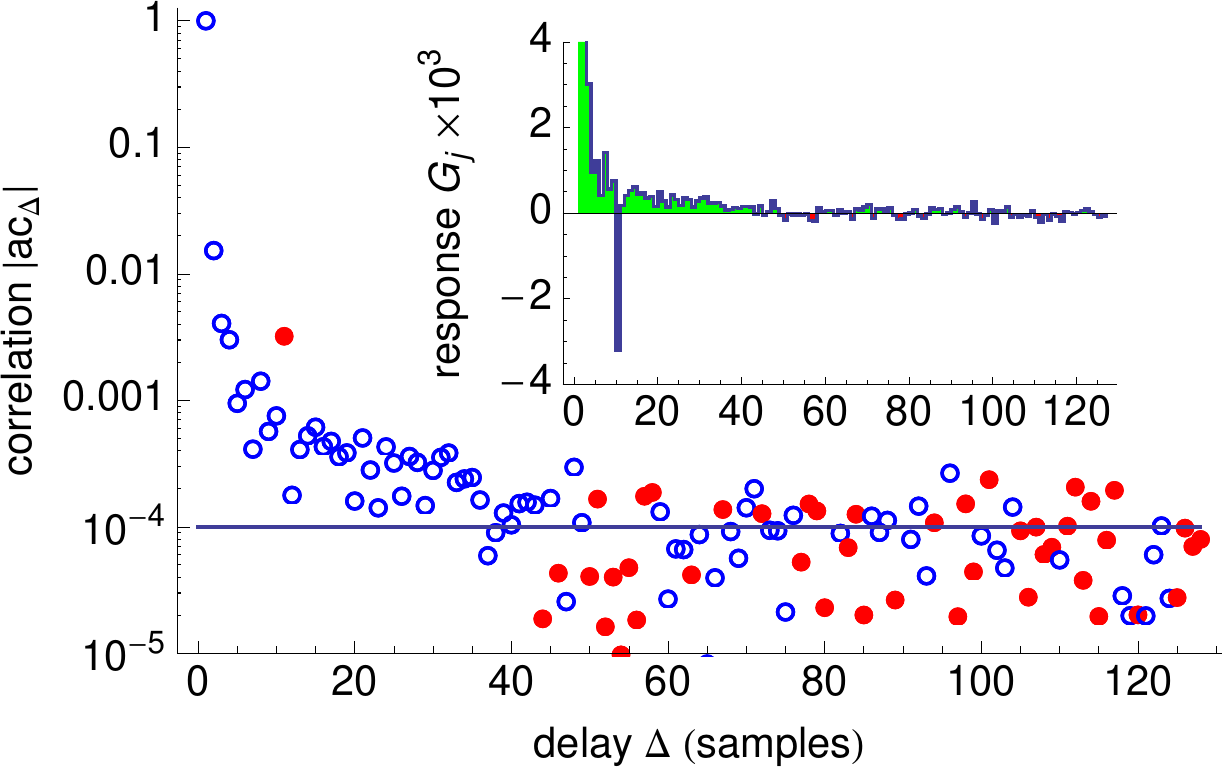}
\caption{(Color online) Normalized correlation and recovered impulse response. The main graph shows autocorrelation $\ac_\acdel$ computed on a string of $10^8$ symbols.  Open blue (solid red) circles indicate positive (negative) correlation.  The horizontal line shows  sampling uncertainty.  The inset shows the reconstructed impulse response function $G_j$, as described in the text.       }
\label{fig:impulse}
\end{figure}

\newcommand{\hang}{\zeta}
\newcommand{\hangover}{hangover}

\section{finite bandwidth}
\label{sec:bandwidth}
Fig. \ref{fig:realtime}~{(middle)} illustrates something intrinsic to analog randomness generators.  An ideal physical process would produce {independent} random values, but this is impossible in a real system due to bandwidth limitations.  When a digital sample is taken, the detection system is still responding (possibly weakly) to analog inputs it received at earlier times.  This is evident in the upper trace of Fig. \ref{fig:realtime}, which visibly {shows electronic ringing and does not fully return to baseline after a strong pulse.}

We model this behavior using Eq. (\ref{eq:V}), but considering only the sampling times $t=t_1, t_2, \ldots$ and write $V_i \equiv V(t_i)$, $G_j \equiv G(t_j)$, etc.,
\bea
V_i &=& \sum_{j=0}^\infty G_j p_{i-j} + \vel_i
\eea
We compute the autocorrelation $\ac_{\acdel} \equiv \cov(V_i,V_{i+\acdel})  = \sum_{jk} G_j G_{k} \cov(p_{i-j}, p_{i+\acdel-k}) = \var(p)  \sum_{j} G_j G_{j+\acdel}$, plus a contribution from $\vel$, and we have assumed $\cov(p_{i},p_{j}) = \var(p) \delta_{ij}$.  For our system, the $\vel$ contribution is negligible:  $\var(V)$ places an upper bound on $\var(\vel)$ for any input power $p$.  Yet, if we interrupt one arm of the interferometer, we observe nearly constant signals $V$, as shown in Fig.~\ref{fig:histos}, with variance 39 dB below the variance of the interference signal.  Because $\ac_\acdel$ can be directly measured from the data, we have an experimental determination of $\ac_\acdel \equiv \sum_{j} G_j G_{j+\acdel}$, the autocorrelation of the impulse response.  Considering that $G_0 \gg G_{j\ne 0}$, and using the causality condition $G_{j<0} = 0$, we find $G_j$ perturbatively as follows.  We write {$G_j \equiv \sum_{n=0}^\infty G_j^{(n)} \lambda^n$}, where $\lambda$ is a parameter that  later is set to unity, and define the cross correlation $\cc_\acdel^{(n,m)} \equiv \sum_{j=0}^\infty G^{(n)}_j G^{(m)}_{j+\acdel}$.  We write 
\bea
\ac_\acdel &=& \lambda^0 ~  \cc_\acdel^{(0,0)} +  \lambda^1 \left[ \cc_\acdel^{(0,1)} + \cc_\acdel^{(1,0)} \right] 
\nnp  \lambda^2 \left[ \cc_\acdel^{(0,2)} + \cc_\acdel^{(1,1)} + \cc_\acdel^{(2,0)} \right] +   \ldots ,
\eea
{and solve by orders in $\lambda$ from the starting condition $G_j^{(0)} \propto \delta_{0,j}$.  Considering the $\lambda^0$ contribution we find $\cc_\acdel^{(0,0)} = [G_0^{(0)}]^2 \delta_{0,\Delta}$ giving the $\lambda^0$ solution $[G_0^{(0)}]^2 = \ac_0$.  Without loss of generality we take $G_0^{(0)}$ to be positive.  Considering then $\lambda^{(1)}$, we solve $\ac_\acdel =  \cc_\acdel^{(0,0)} + \left[ \cc_\acdel^{(0,1)} + \cc_\acdel^{(1,0)} \right]$, a linear equation for $G_j^{(1)}$,  by matrix inversion.  Continuing in a similar fashion for higher orders in $\lambda$, $G_j$ rapidly converges to give the impulse response shown in Fig. \ref{fig:impulse}.  Considering the low degree of observed correlation, it is not surprising that this  resembles the correlation $\ac_\acdel$ and is dominated by the $\Delta=0$ term.}  It is perhaps interesting to note the narrow negative feature at $\acdel=10$, probably due to an electronic reflection in the cabling of the digitization electronics.  

The net contribution of previous pulses is $\vprev_i = \sum_{j=-\infty}^{i-1} p_j G_{i-j}$.  This contributes to the variance of individual  $V_i$ without adding any randomness to the sequence.  From the $d_i$ sequence we find bounds $\hang_- \equiv \min_i \vprev_i  = -0.0145 $ full scale, or -3.7 {codes} at 8-bit resolution, $\hang_+ \equiv \max_i \vprev_i  = 0.0156 $ full scale, or +4.0 {codes} at 8-bit resolution.  We refer to $\hang_-$ and $\hang_+$ as ``\hangover~ errors'' for their delayed nature. 

\section{Refinement of the problem}

Having established a model for the device, we now ask the following: Trusting only $\phiq$ to be random, how much randomness exists in the output string ?  In particular, we do not trust $\uone,\utwo,\vis,\phic,\vel$, or $\error$ to be random.  Fluctuations in these quantities can be traced to fluctuations of classical variables, for example, the injection current of the diode, that certainly contain patterns, and that could, in principle, be described by a perfectly deterministic pattern unknown to us.  We are not, however, completely ignorant about these quantities; their distributions are constrained by the digitization and correlation measurements described above, as well as  by the distributions  of  $\di$,  
$\ds$, and $\dl$.

A key observation is illustrated by Fig. \ref{fig:histos}~{(bottom)}.   The distributions of $\ds$ and $\dl$ are very narrow, whereas the distribution of $\di$ is broad.  Provided the digitization gives a not-too-unfaithful conversion from $\u$ to $d$, we conclude that $\udet$ varies much more than $\uone$ or $\utwo$.  By Eq. (\ref{eq:interference}), this implies $\vis \ne 0$, at least for some fraction of the measured pulses.  $\vis \ne 0$ in turn means that $\udet$ (and thus $\di$) contain some randomness from $\phiq$.    Our  goal is to make quantitative this observation, to put lower bounds on the quantum randomness of the string  $\{\di_i\}$.

\section{digitization limits}
\label{sec:digilimits}
An ideal digitization process would output the value $d \in [0,N-1]$ for inputs in the range $p \in [p_{d,-}^{(\rm ideal)},p_{d,+}^{(\rm ideal)})$, where 
\begin{eqnarray}
\label{eq:CDF}
p_{d,-}^{(\rm ideal)} & \equiv & \left\{ 
\begin{array}{rl}  
-\infty & d=0 \\
d & {\rm otherwise}
\end{array}
\right. \\
p_{d,+}^{(\rm ideal)} & \equiv & \left\{ 
\begin{array}{rl}  
\infty & d = N-1 \\
d+1 &  {\rm otherwise}
\end{array}
\right. 
\end{eqnarray}

We have seen, however, that our digitizer sometimes makes errors; i.e. it outputs a value $d$ when $p \notin [p_{d,-}^{(\rm ideal)},p_{d,+}^{(\rm ideal)})$.  The distribution of these errors is illustrated in Fig. \ref{fig:digilimits}, and can be roughly characterized by the rms width $\approx 0.8$ {codes.}  Defining $p_{d,-}^{(\rm dig)}$ and $p_{d,+}^{(\rm dig)}$ as the minimum and maximum inputs, respectively, that are seen to give rise to an output $d$, we can say with confidence that an output $d$ implies an input $p \in [p_{d,-}^{(\rm dig)},p_{d,+}^{(\rm dig)})$.  This also allows us to bound the probability $P(d)$  of an output $d$.  Given a cumulative distribution function (CDF) $F(p)$ for the input,  the output satisfies 
 $P(d) \le  F(p_{d,+}^{(\rm dig)}) - F(p_{d,-}^{(\rm dig)})$.

We can include also errors due to finite bandwidth in this description.  If the minimum and maximum \hangover~are $\hang_-$ and $\hang_+$, respectively (cf. Sec. \ref{sec:bandwidth}), then a value $d$ implies $p \in [p_{d,-}^{(\rm d+ h)},p_{d,+}^{(\rm d+ h)})$, where $p_{d,\pm}^{(\rm d+ h)} = p_{d,\pm}^{(\rm dig)} + \hang_\pm$ (the superscript $^{(\rm d+h)}$ indicates the combined effects of digitization and hangover errors).  These digitization limits including hangover will be used to evaluate digitization of the strongly varying signal $\udet$, while the limits without hangover will be used for the weakly varying $\uone$ and $\utwo$, for which the hangover error is negligible.

\begin{figure}[tr]
\includegraphics[width=0.175 \textwidth]{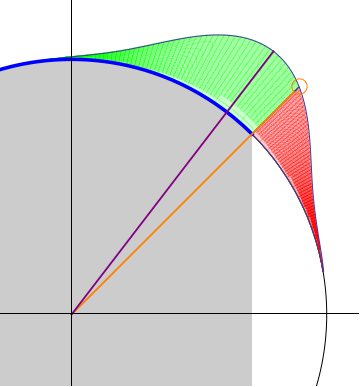}
\includegraphics[width=0.3 \textwidth]{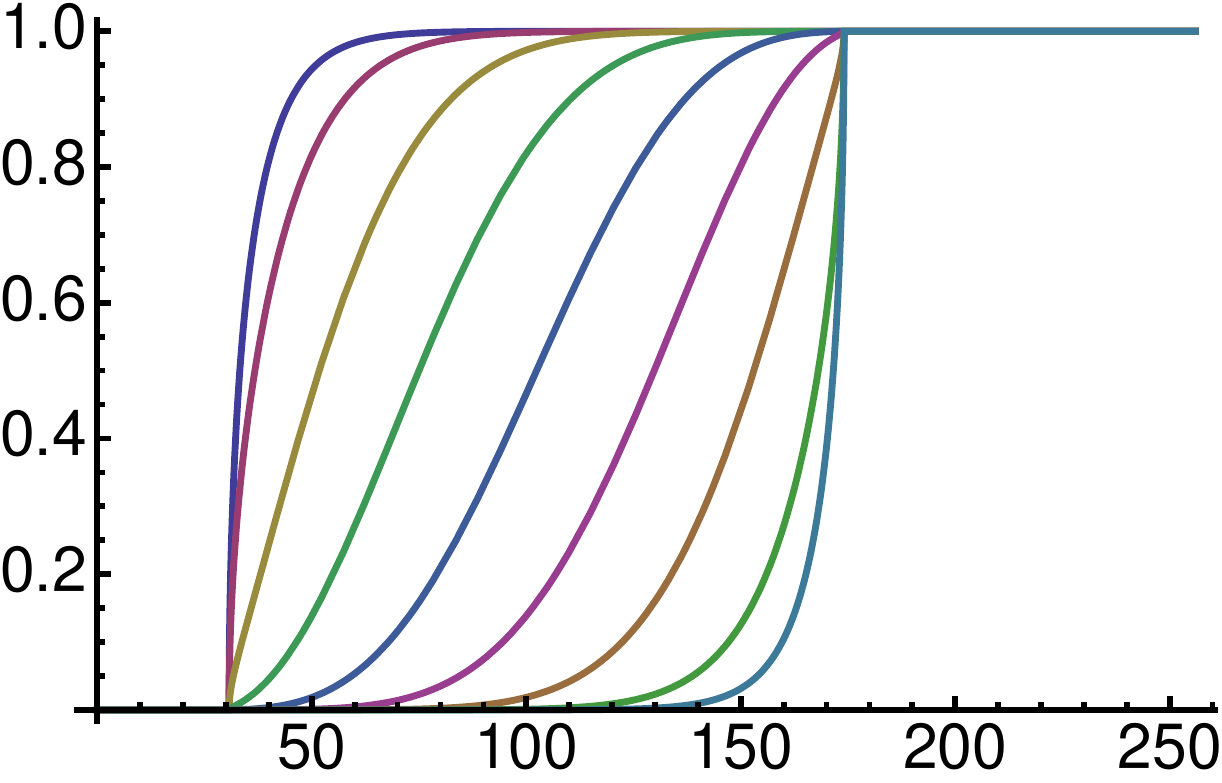}
\caption{(Color online) Illustration of the distribution function $\CDF_{\sigq}(\udet|\xvec)$ that characterizes $\udet$ given by Eq. (\ref{eq:interference}) for fixed $\uone, \utwo, \phic$, and normally distributed $\phiq$.  (Left) Visualization of the calculation.  Gaussian $P(\phiq)$ (radial coordinate) centered at $\phic$ (polar coordinate), has probability mass {(green area)} given by the error function between limits given by the arccosine of the scaled and shifted $\udet$ (horizontal coordinate).  (Right) Illustration of  $\CDF_{\sigq}(\udet|\xvec)$ for $\uone = \utwo = 51$, $\vis = 0.7$,  $\sigq = \pi/8$, and  $\phic = 0, \pi/8, \ldots , \pi$, from left to right.  
}
\label{fig:CDF}
\end{figure}

\section{possible distributions}

For given $\xvec \equiv (\uone,\utwo,\vis,\phic)$, and with $\phiq$  normally distributed with mean zero and rms width $\sigq$, we can compute $\CDF_{\sigq}(\udet| \xvec)$, the CDF for $\udet$, as follows.  We note the transformation of variables rule: If $Y = f(X)$, where $f$ is a differentiable function  and $X$ is a random variable with distribution $P_X(X)$, then the distribution of $Y$ is
\begin{equation}
\label{eq:DistributionOfAFunction}
P_Y(Y) = \sum_i \left| \frac{d}{dY} f_i^{-1}(Y) \right| P_X(f_i^{-1}(Y))
\end{equation}
where $f_i^{-1}(Y)$ indicates the $i$'th root of the equation $f(X) = Y$.   Applied to Eq. (\ref{eq:interference}) and integrating to find $\CDF_{\sigq}(\udet| \xvec)$ from $P_{\udet}(\udet)$, we find 
\begin{eqnarray}
\label{eq:CDF}
CDF_{\sigq}(\udet| \xvec) &=& 1- \frac{1}{2}
\sum_{n=-\infty}^{\infty} \left. \erf \frac{\phi -\phic + 2 \pi n}{\sigq \sqrt{2}}\right|_{\phi=-\phidet}^{\phi=\phidet}, \,\, \\
\phidet & \equiv & \arccos  \frac{ \udet- \uone - \utwo  }{2\vis \sqrt{\uone \utwo} }
\end{eqnarray}
where $\erf$ is the error function. This result is illustrated in Fig. \ref{fig:CDF}.  
The CDF has the usual interpretation: The probability to find $\udet$ in an interval $[a,b)$ is $\CDF_{\sigq}(b|\xvec) - \CDF_{\sigq}(a|\xvec) $.  

We are also  interested in the case where $\phiq + \phic$ is completely uncertain, or equivalently uniformly distributed on $[0,2\pi)$.  This gives
\begin{eqnarray}
\label{eq:CDFu}
\CDF_{\circ}(\udet|\xvec) 
& \equiv &  1-  \frac{1}{\pi} {\rm Re}[ \arccos  \frac{ \udet- \uone - \utwo   }{2\vis \sqrt{\uone \utwo} } ],
\end{eqnarray}
which not surprisingly is the \mbox{$\sigq\rightarrow \infty$} limit of $\CDF_{\sigq}(\udet| \xvec)$. 

Finally, for the non-interfering signals $\uone$ and $\utwo$, the relevant CDF is 
\begin{eqnarray}
\label{eq:Fuoneutwo}
\CDF_{|,s}(p|\xvec)  &\equiv&  \theta(p-p^{(\rm s)}), \\
\CDF_{|,l}(p|\xvec)  &\equiv&  \theta(p-p^{(\rm l)}),
\end{eqnarray}
where $\theta$ is the Heaviside step function.   Given a CDF $F(P|\xvec)$ and a distribution $P(\xvec)$ for $\xvec$, the statistically averaged CDF is 
\begin{eqnarray}
\label{eq:FfromFofX}
F(p) = \int d^4 \xvec \, \CDF(p|\xvec) P(\xvec).
\end{eqnarray}

\newcommand{\comberr}{\epsilon_{\rm comb}}

The $\udet$ digitization frequencies of Fig. \ref{fig:histos} were collected with $\phic$ varying due to thermal expansion of the fiber loop in the MZI, and probably several other factors.  This causes a drift by much more than $2 \pi$ over the time of the acquisition, so it is appropriate to compare the $\udet$ data against $\CDF_{\circ}(\udet)$, which incorporates the $\phic$ averaging.   
{If we write $P^{(\rm s)}(d)$, $P^{(\rm l)}(d)$, and $P^{(\rm i)}(d)$ for the probabilities of digitization outcome $d$ when measuring variable $p^{(\rm s)}$, $p^{(\rm l)}$, and $p^{(\rm i)}$, respectively, then the probability of an outcome in the range $l$ to $h$ is $P_{l,h}^{(\rm  s)} \equiv \sum_{d=l}^h P^{(\rm s)}(d)$ and similar for $P_{l,h}^{(\rm  l)}$ and $P_{l,h}^{(\rm  i)}$.} $P_{l,h}^{(\rm  i)}$ is upper bounded by 
\begin{eqnarray}
\label{eq:PBounds1}
P_{l,h}^{(\rm i)}& \le &   \CDF_{\circ}(p_{h,+}^{(\rm d+h)} ) - \CDF_{\circ}(p_{l,-}^{(\rm d+h)} ),
\end{eqnarray}
where $[p_{d,-}^{(\rm d+h)}, p_{d,+}^{(\rm d+h)} )$ is the range, including errors as described above, of the digitization outcome $d$.   
We can also obtain a lower bound, considering that $P_{l,h} = 1- P_{0,l-1} - P_{h+1,N-1}$, and that the latter two terms are upper bounded as above.  We find 
\begin{eqnarray}
\label{eq:PBounds2}
P_{l,h}^{(\rm i)}& \ge &  \CDF_{\circ}(p_{h+1,-}^{(\rm d+h)} ) - \CDF_{\circ}(p_{l-1,+}^{(\rm d+h)} ) .
\end{eqnarray}
As both  $P(d)^{(\rm i)}$and the limits $p_{d,-}, p_{d,+}$ have been measured, Eqs. (\ref{eq:PBounds1}) and (\ref{eq:PBounds2})  provide experimental constraints on $P(\xvec)$.  

Analogous constraints apply to the noninterfering signals
\begin{eqnarray}
\label{eq:PBounds3}
P_{l,h}^{(\rm s)} & \le &   \CDF_{|,s}(p_{h,+}^{(\rm dig)} ) - \CDF_{|,s}(p_{l,-}^{(\rm dig)} )  \\
\label{eq:PBoundsLast}
P_{l,h}^{(\rm s)} & \ge &  \CDF_{|,s}(p_{h+1,-}^{(\rm dig)} ) - \CDF_{|,s}(p_{l-1,+}^{(\rm dig)} )
\end{eqnarray}
and similar for $P_{l,h}^{(\rm l)}$.

\newcommand{\worst}{{\rm wc}}

\section{randomness quantification Redux}

{We now find a lower bound for $\AME$, as in Sec. \ref{sec:RandomnessQuantification1}, but including worst-case considerations for digitization and hangover errors.   }
As above, we first consider a given $\xvec$, implying a given $\CDF_{\sigq}(\udet| \xvec)$.  Inclusion of digitization and correlation errors leads to the upper bound
\begin{eqnarray}
\label{eq:PBoundsH}
P^{(\rm i)}(d|\xvec)& \le &  F_{\sigq}(p_{d,-}|\xvec ) - F_{\sigq}(p_{d,+}|\xvec ).
\end{eqnarray}
In contrast to $\uone$, $\utwo$, and $\vis$, which are more-or-less directly reflected in $\{ d_i \}$ and thus have distributions constrained by, e.g., Eq. (\ref{eq:PBounds1}), we have little measured  information about $\phic$.  To be conservative, we maximize the right-hand side over this variable to find the ``worst-case''  (wc) bounds
\begin{eqnarray}
\label{eq:PBoundsH}
P^{(\rm i)}(d|\xvec)& \le &   \max_{\phic} \left[ F_{\sigq}(p_{d,-}|\xvec ) - F_{\sigq}(p_{d,+}|\xvec ) \right]  \nnequiv P^{(\worst)}(d|\xvec). \, \,
\end{eqnarray}
{Now $\max_d  P^{(\worst)}(d|\xvec)$ upper bounds the predictability of a single symbol, produced with a given $\xvec$.  
For a string of symbols, generated as $\xvec$ varies with distribution $P(\xvec)$, the average min-entropy is lower bounded by Eq.~(\ref{eq:AMEdef}) applied to $P^{(\worst)}(d|\xvec)$: 
\begin{equation}
\AMEd \ge - \log_2 \int d\xvec \, P(\xvec)  \max_d P^{(\worst)}(d|\xvec) \equiv  \AME^{(\worst,P(\xvec))}.
\end{equation}
}

\begin{figure}[tr]
\includegraphics[width=0.22 \textwidth]{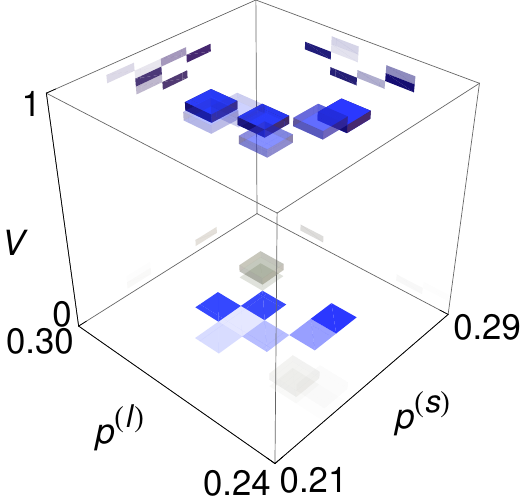} ~ ~
\includegraphics[width=0.22 \textwidth]{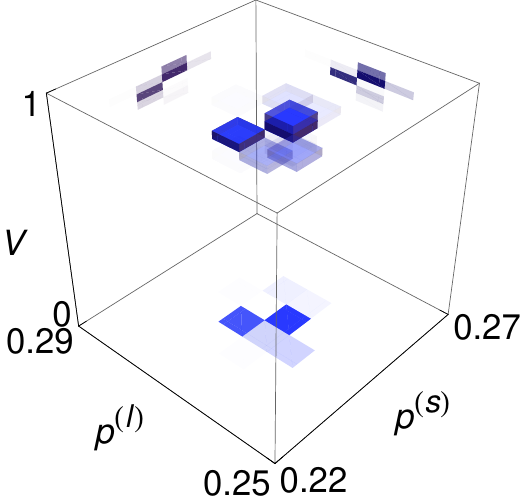}

\caption{(Color online) Optimized piecewise-constant distribution $P(\xvec)$ for 8-bit digitization and $\sigq = 3\pi/2$.  Axes indicate $\uone$, $\utwo$, and $\vis$; density indicates $s_i$.  $\phic$ is not included as an independent dimension because it is chosen according to other criteria (see text).  
 The ranges of $\uone$ and $\utwo$ are chosen to
 cover the whole range of these variables allowed by the measured distributions  shown in Fig. \ref{fig:histos}, in light of digitization errors from Fig. \ref{fig:digilimits}. The graphic on the left uses  worst-case errors (green curves in Fig. \ref{fig:digilimits}); the one on the right uses error limits narrower by a factor $0.275$. 
 Within these ranges, the space is divided into a uniform $8\times 8 \times 32$ rectangular grid $\{ \xi_i \}$, and corresponding weights $\{ s_i \}$ are calculated by numerical minimization of the min-entropy lower bound as in Sec. \ref{sec:optimization}.  The probability is concentrated in regions of high visibility, necessary to agree with the wide measured distribution, and regions of low visibility, which give low min-entropy.  The distributions of $\udet, \uone$, and $\utwo$ that follow from these $P(\xvec)$ are shown in Fig. \ref{fig:DataAndPred}.  }
\label{fig:OptimizedDistribution}
\end{figure}

\begin{figure}[tr]
\includegraphics[width=0.45 \textwidth]{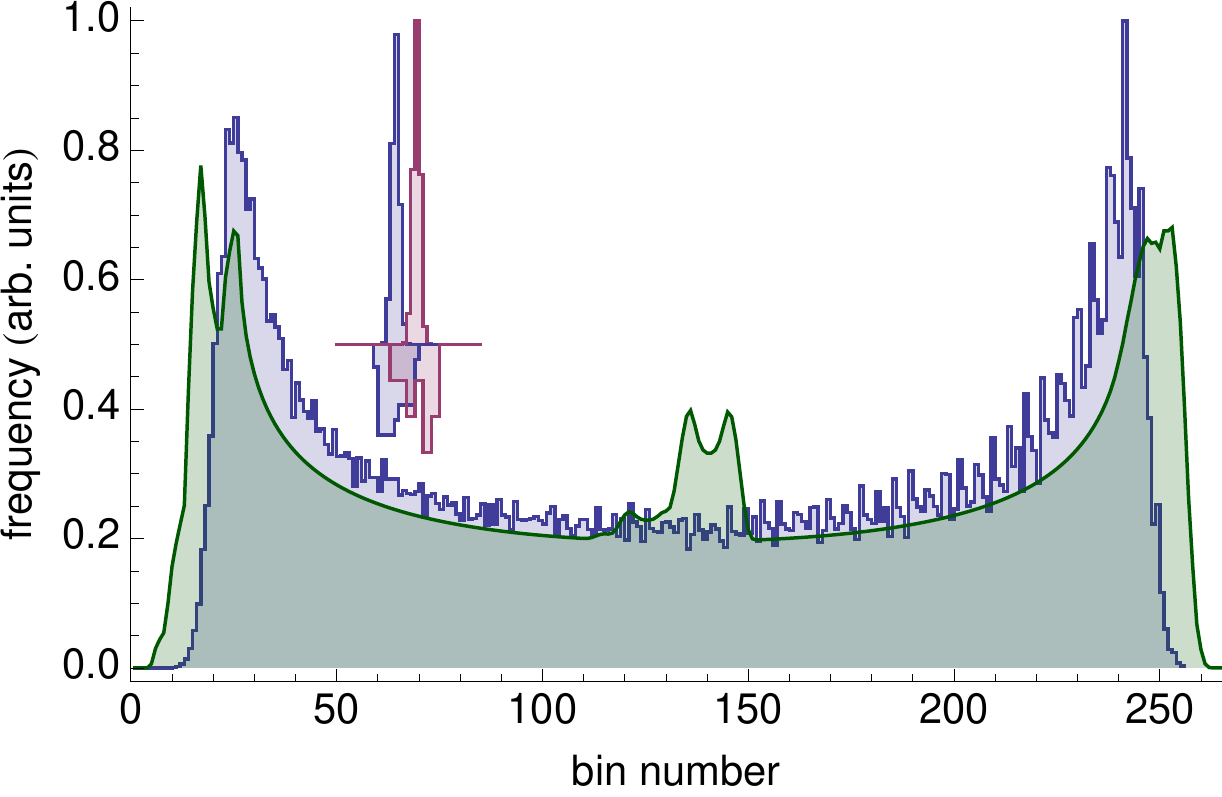}
\includegraphics[width=0.45 \textwidth]{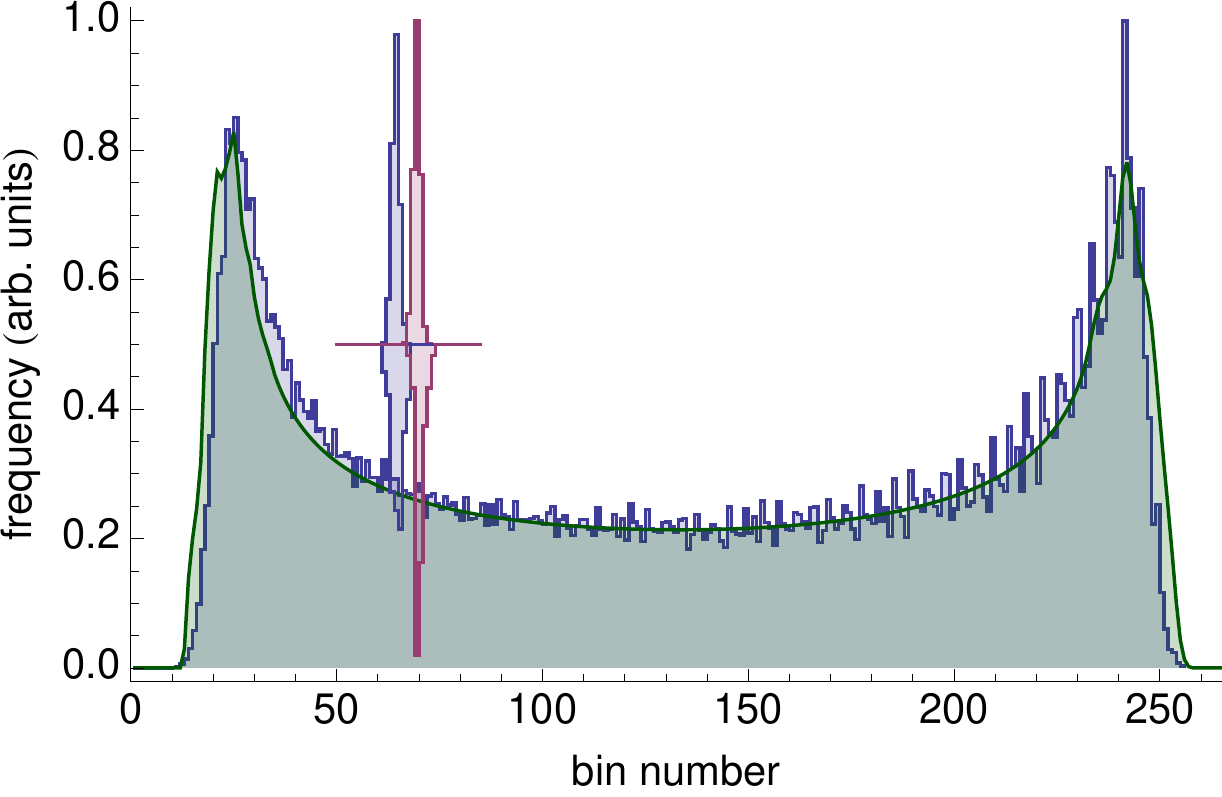}
\caption{(Color online) Comparison of measured frequencies against their most conservative interpretation, the prediction from the optimized $P(\xvec)$.  (Top) Prediction from $P(\xvec)$ of Fig.~\ref{fig:OptimizedDistribution} (left), assuming worst-case tolerances.  (Bottom) Prediction from $P(\xvec)$ of Fig.~\ref{fig:OptimizedDistribution} (right), assuming tolerances 0.275 of worst case.   The main graph shows a histogram of observed $\udet$ (jagged blue), and vertically offset $\uone, \utwo$ (inset, left blue and right red), the same as in Fig. \ref{fig:histos}.  Superposed smooth green curves show the predicted distribution $P(\udet)$ computed from $P(\xvec)$ chosen to minimize $\AME^{(\worst, \{ \chi_i \})}$.  The inset shows, inverted, the predicted distributions for $\uone$ and $\utwo$.  The predicted distributions are consistent with the observed data in light of the tolerances provided by digitization and hangover errors (see Secs. \ref{sec:digitization} and \ref{sec:bandwidth}).  Note the central bump, from to low-visibility parts of the distribution, that lowers the  min-entropy. }
\label{fig:DataAndPred}
\end{figure}

\begin{figure}[tr]
\includegraphics[width=0.45 \textwidth]{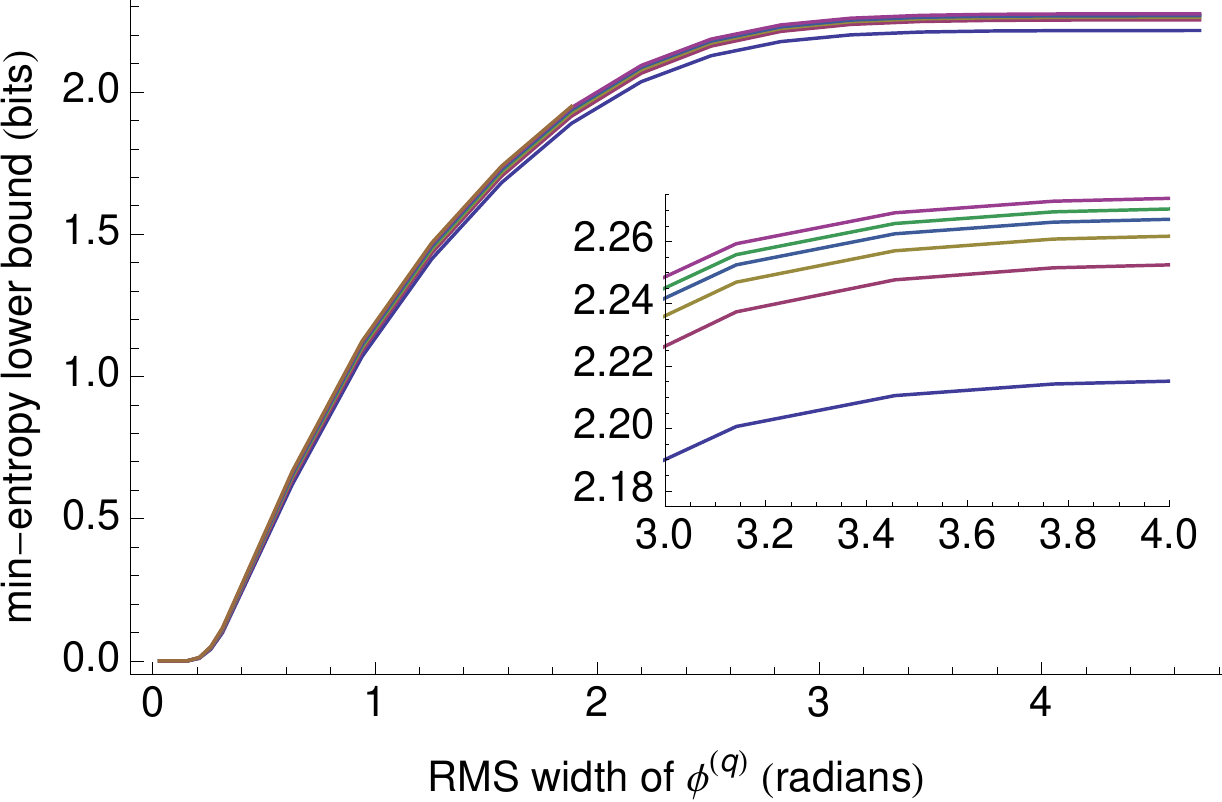}
\caption{(Color online) Min-entropy bound as a function of $\sigq$ for rectangular-lattice coverings of different resolution.   Digitization is 8 bits.  With $n(1\times 1 \times 4)$ divisions, where $n = 6, \ldots ,12$,  giving the shown curves, from bottom 
to top.   The inset shows the same curves on a finer scale.  Increasing $n$ gives an increasing lower-bound for $\AME$: The inevitable error due to finite covering resolution works to reduce $\AME$, making the estimate conservative.  With $n= 8$ we find $1\%$ accuracy relative to $n=12$, the highest resolution we could optimize using the MATLAB function {\tt linprog} and 8 GB of RAM.  
}
\label{fig:MinEntVsSig}
\end{figure}

\section{optimization}
\label{sec:optimization}

{Our goal is now to minimize $ \AME^{(\worst,P(\xvec))}$, or equivalently to maximize 
\begin{equation}
\label{eq:MeanPredWCDef}
\meanPredwc \equiv \int d\xvec \, P(\xvec)  \max_d P^{(\worst)}(d|\xvec)
 \end{equation}
 }
by choice of $P(\xvec)$, subject to constraints as in Eqs. (\ref{eq:PBounds1})--(\ref{eq:PBoundsLast}).  This will give a conservative estimate of contribution of $\phiq$ to the min-entropy in the digitized bit string.  We transform this into a linear programming problem by splitting the $\xvec$ space into a covering by nonoverlapping regions $\{ \chi_i \}$. {If $R_{\chi_i}(\xvec) \equiv 1$ for $\xvec \in \chi_i$ and zero otherwise, then 
the probability to find  \mbox{$\xvec\in\chi_i$} is  $s_i \equiv \int d^4  \xvec \, R_{\chi_i}(\xvec) P(\xvec)$. 
By assumption $\int d^4  \xvec \, R_{\chi_i}(\xvec) R_{\chi_j}(\xvec)=0$ for $i\ne j$. 

Inserting the identity $\sum_i R_{\chi_i}(\xvec) $ in Eq. (\ref{eq:MeanPredWCDef}) we find
\begin{eqnarray}
\label{eq:PBoundsH}
 \meanPredwc 
& = & \int d^4 \xvec \, \sum_i R_{\chi_i}(\xvec)  P(\xvec)  \max_d P^{(\worst)}(d|\xvec) \, \,\, \,  \\
\label{eq:PBoundsH3}
& \le &  \sum_i \int d^4 \xvec \, R_{\chi_i}(\xvec)  P(\xvec)  \max_{\xvec \in \chi_i}  \max_d P^{(\worst)}(d|\xvec)  \\
\label{eq:PBoundsH3b}
& = & \sum_i  s_i \max_{\xvec \in \chi_i} \max_d P^{(\worst)}(d|\xvec)  \\
\label{eq:PBoundsH4}
& \equiv & \meanPred^{(\worst,\{ s_i , \chi_i \})}
\end{eqnarray}
As described below, the maximization over ${\xvec \in \chi_i}$ in Eq. (\ref{eq:PBoundsH3}) makes the coarse-graining procedure conservative.
}

{
The probabilities $s_i$ are constrained by $\int d^4 \xvec \, P(\xvec)  = 1$ or
\begin{eqnarray}
\label{eq:SConstraint}
 \sum_i s_i  & = &  1.
\end{eqnarray} 

An additional set of constraints, also linear in the $\{ s_i \}$, is generated from Eqs. (\ref{eq:PBounds1})--(\ref{eq:PBoundsLast}) by applying the coarse-grained average 
\begin{equation}
\label{eq:CG}
P(\xvec) \rightarrow \sum_i s_i  R_{\chi_i}(\xvec),
\end{equation} 
to Eq. (\ref{eq:FfromFofX}), to give 
\begin{eqnarray}
\label{eq:FfromFofXcg}
F(p) \rightarrow \sum_i s_i \int d^4 \xvec \, \CDF(p|\xvec) R_{\chi_i}(\xvec),
\end{eqnarray}
describing the various $F$ quantities appearing in Eqs. (\ref{eq:PBounds1})--(\ref{eq:PBoundsLast}).
 In what follows, the $\chi_i$ are chosen to be rectangular regions of $\xvec$ space, which facilitates the necessary integrations.  For example, $\int d\vis \, \CDF_{\circ}(\udet|\xvec)$ has an analytic form, reducing the number of numerical integrals. 

}

Having expressed the constraints and objective function as linear functions of the $s_i$, we use a large-scale linear programming routine to find the unique solution {$\{ s_i \}$} that { maximizes $ \meanPred^{(\worst,\{ s_i , \chi_i \})}$} subject to the set of constraints, for a given covering $\{ \chi_i \}$.   {We arrive to the bound  
\begin{equation}
\AMEd \ge - \log_2 \max_{\{s_i \}} \meanPred^{(\worst,\{ s_i , \chi_i \})} \equiv \AME^{(\worst, \{ \chi_i \})} .
\end{equation}} Illustrations are given in Figs.  \ref{fig:DataAndPred} and \ref{fig:OptimizedDistribution}.   We increase the resolution, i.e., increase the number of elements in the covering while decreasing their volumes, to reach our best estimate of  {$\AME^{(\worst, \{ \chi_i \})}$}.  { Because the target function {$\AME^{(\worst, \{ \chi_i \})}$} is calculated using the worst point in each region, as in Eq. (\ref{eq:PBoundsH3}), while the constraints are calculated using the region average, as in Eq. (\ref{eq:CG}),} the average min-entropy bound  increases with increasing resolution, making the procedure conservative at finite resolution. See Fig.   \ref{fig:MinEntVsSig} for illustration.  

{
%
%
The statistical analysis described here can, in principle, be performed on the raw data themselves, i.e., applied to the symbols $\{ d\}$ prior to randomness extraction.  Furthermore, the analysis uses only the frequencies of the symbols and is independent of their order.  For these reasons,  there is no reason $P(\xvec)$ must be stationary in time.  Rather, it describes the distribution of $\xvec$ aggregated over the time of the data acquisition.  }

\begin{figure}[tr]
\includegraphics[width=0.45 \textwidth]{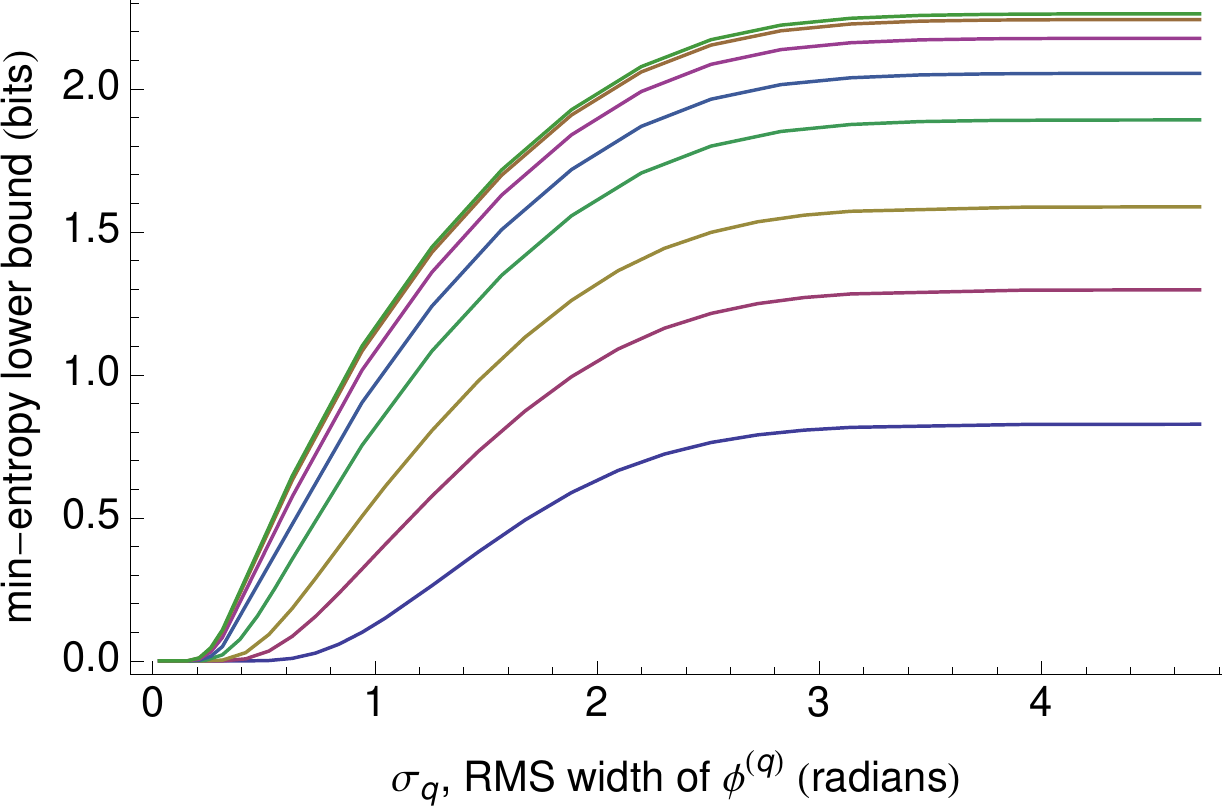}
\caption{(Color online) Lower bound on min-entropy versus $\sigq$  for different digitization resolution, from 1 bit to 8 bit (bottom to top).  Other conditions are:  covering resolution {$(\uone,\utwo,\vis) = 8 \times 8 \times 32$}, ``worst-case'' assumptions for digitization and hangover errors.}
\label{fig:MinEntVsSigBitDepth}
\end{figure}

\begin{figure}[tr]
\includegraphics[width=0.45 \textwidth]{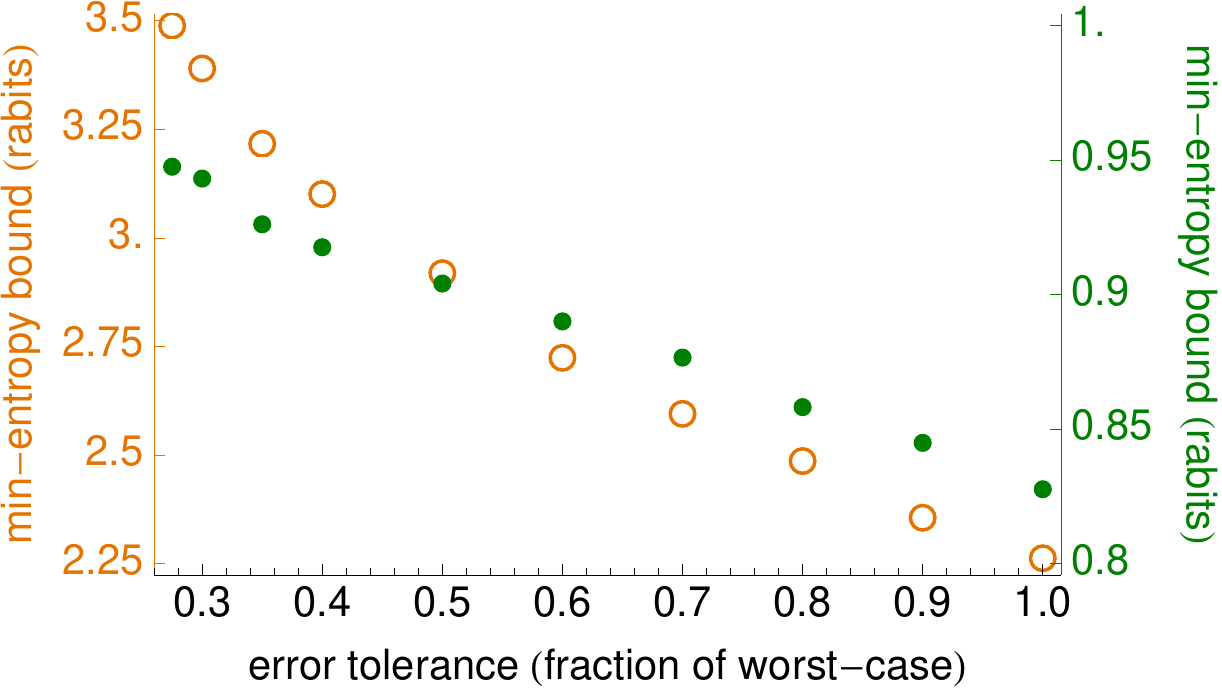}
\caption{(Color online) Lower bound on min-entropy versus error tolerance at $\sigq = 3 \pi/2$ and covering resolution $(\uone,\utwo,\vis) = 8 \times 8 \times 32$.  Hollow orange circles show 8-bit digitization (on left scale), filled green circles show binary digitization (on right scale).  Error limits for a given $d$ are computed using the data shown in Fig. \ref{fig:digilimits}, plus the hangover errors $\zeta_\pm$ for $\udet$ digitization, and are interpolated between the mean and the worst-case limits by the error tolerance 
shown here on the horizontal axis.  For 
error tolerance
below $0.275$ and with this covering, no $P(\xvec)$ is consistent with the distributions shown in Fig. \ref{fig:histos}.   }
\label{fig:MinEntVsErrorTolerance}
\end{figure}

\section{experimental results  }

{We apply the above analysis to the QRNG described in \cite{AbellanOE2014}, based on the data shown in Figs. \ref{fig:histos}, \ref{fig:digilimits}, and \ref{fig:impulse}.  To apply the analysis, we need a value for $\sigma_q$, which we take to be $\sigma_q = 3 \pi/2$, well into the plateaux seen in Fig. \ref{fig:MinEntVsSigBitDepth}. Previous works describing the same system \cite{Jofre2011,AbellanOE2014} describe a rapid phase diffusion, reaching $\sigma_q > 3 \pi$ after a diffusion time of 0.17 ns.  Our 5-ns diffusion time is 29 times longer, and thus $\sigma_q = 3 \pi/2$ is very conservative.  The results of \cite{AbellanOE2014} are based on modeling of the laser dynamics (see also the Appendix), supported by direct experimental observations of the pulses.   By considering systematic uncertainties in the laser parameters, and statistical uncertainties in the observations, it would, in principle, be possible to place a confidence level on the assertion that $\sigma_q \ge 3 \pi/2$.   In this case, however, we can see no reasonable scenario in which the phase diffusion is  so much slower (at least a factor of  $58$) than calculated; the experimental results of \cite{AbellanOE2014} would have been dramatically different in that case.  
}

Fig. \ref{fig:MinEntVsSigBitDepth}  shows {$\AME^{(\worst, \{ \chi_i \})}$, the lower bound on the average min-entropy,} as a function of digitization resolution.  We find a lower bound of 2.3 quantum random bits  per symbol with 8-bit digitizationy-, and 0.83 quantum random bits per symbol with binary digitization.  Constraints are computed as above, from the 8-bit characterization measurements, and we compute lower-resolution digitizations by splitting the range $\udet \in [0,1)$ into $N=2^b$ equally spaced bins.  We assume worst-case digitization and hangover errors as in Sec. \ref{sec:digilimits}.  The results show a roughly linear increase in {$\AME^{(\worst, \{ \chi_i \})}$} versus $b$ until saturation around $b=6$.  This supports the intuitively reasonable conclusion that resolution finer than the scale of the digitization errors contributes little to {$\AME^{(\worst, \{ \chi_i \})}$}.  

The above results are obtained with a  high degree of statistical confidence.  As described in Sec. \ref{sec:digitization}, we use as our error limits the most extreme errors seen in $2^{14}$ samplings for any given digitization output.  We thus have a confidence level of $1-2^{-14} \approx 0.999939 $ that any given digitization event will be within our limits and thus is properly accounted for in computing the average min-entropy.  For hangover errors, due to a larger data set, this confidence is $\sim 1-10^{-8}$.   It will surely be reasonable to consider less conservative error bounds for some applications.  {We define a fractional error tolerance $\eta$ as follows:  Recall that  $p_{d,\pm}^{(\rm ideal)}$ and $p_{d,\pm}^{(\rm d+ h)}$ are the minimum ($-$) and maximum ($+$) values that can give rise to a symbol $d$ in the ideal and error-adjusted cases, respectively.   Corresponding limits with scaled errors are $p_{d,\pm}^{({\rm d+ h},\eta)} \equiv \eta p_{d,\pm}^{(\rm d+ h)} + (1-\eta) p_{d,\pm}^{(\rm ideal)}$.  
In Fig. \ref{fig:MinEntVsErrorTolerance} we show $\AME^{(\worst, \{ \chi_i \})}$ versus $\sigq$ for different   $\eta$, showing up to {3.5} quantum random bits per symbol in 8-bit digitization, and up to {0.947} quantum random bits per symbol for binary digitization.  }

\section{conclusions}
Establishing the randomness of data generated by a physical process is a vexing challenge, with important consequences for data security and stochastic simulations.  While many experiments have generated data that in some way reflected the randomness of quantum physics,  many applications require both full randomness and realistic assurances of randomness.  {We have described a  methodology and experimental standard of  proof for quantum randomness, similar to the methodology of precision measurement.   

The methodology is paranoid in the sense that it assumes the worst case behavior for all untrusted variables.  As in precision measurement, it is possible to place experimental constraints on the behavior of these variables using auxiliary measurements and the generated data themselves.  A constrained numerical optimization of the distribution of untrusted variables  gives a lower bound for the average min-entropy, the measure of randomness appropriate to randomness extraction.  This enables the generation of nearly perfect $\epsilon$-random bit strings.  A confidence level, also paranoid,  is assigned to the average min-entropy estimate, and thus to the $\epsilon$-randomness of the generated string.  

We apply the method to an ultrafast phase-diffusion QRNG, and find the system is an efficient randomness generator even under this paranoid analysis.
The result shows that strong experimental guarantees can be given for quantum random number generators.  
}

\section{acknowledgements}
We thank V. Pruneri for laboratory space, K. Shalm for
helping to define the problem, S. Wehner for advice on
probability theory, and C. Fleischer and K. Turner for careful
readings. This work was supported by the ERC under Project 
AQUMET,  MINECO Project FIS2011-23520, and Fundaci\'{o} 
Privada CELLEX.

~

{
\appendix
\section{Phase diffusion in diode lasers}
\label{app:PhaseDiffusion}

The dynamics of a diode laser are described by a set of stochastic differential equations that govern the 
exchange of energy between the charge carriers (electrons) and the field, driven by the injection current $I$, with noise added from spontaneous emission and spontaneous loss of electrons.  
 We reproduce the equations from Agrawal  \cite{Agrawal1990}.  Other formulations  \cite{ScullyZubairy1997} have similar global properties:
 \begin{eqnarray}
 \label{eq:Pdot}
 \dot{P} &=& (G_L/\sqrt{1+ p} - \gamma) P + R_{\rm sp} + F_P(t) \\
  \label{eq:phidot}
\dot{\phi} &=& \frac{\alpha}{2}(G_L - \gamma)  + \frac{\beta}{2} \frac{G_L p}{1 + \sqrt{1+p}} + F_\phi(t) \\
 \label{eq:Ndot}
\dot{N} &=& I/q - \gamma_e N - G_L P/\sqrt{1+p} + F_N(t).
\end{eqnarray}
Here $P$ is the number of photons, $\phi$ is the phase of the intra-cavity field and $N$ is the number of charge carriers.  $F_P(t), F_\phi(t)$, and $F_N(t)$ are $\delta$-correlated zero-mean Langevin noise terms, giving diffusion coefficients
\begin{equation}
\begin{array}{lll}
D_{PP} = R_{\rm sp}, & ~ D_{\phi\phi} = R_{\rm sp}/(4 P),  & ~ D_{P\phi} = 0  \\
D_{NN} = R_{\rm sp} P + \gamma_e N, & ~ D_{PN} = -R_{\rm sp} P, & ~ D_{N\phi} = 0. 
\end{array}
\end{equation}
Here $R_{\rm sp}$ is the rate of spontaneous emission, which depends on $N$, while $\gamma_e$ is the decay rate of the carrier population.  The other variables describe laser characteristics that are not important in this discussion.  Note that all of the noise terms are traceable to two spontaneous processes: the spontaneous emission of photons $R_{\rm sp}$ and the spontaneous loss of carriers $\gamma_e N$, both of which give rise to $\delta$-correlated noise.  The dynamics are invariant under a global change of $\phi$.

If we write the dynamical equation for $\phi$ as $\dot{\phi} = A + F_{\phi}(t)$, we can formally integrate to find  $\Delta \phi$, the change in $\phi$ over one pulse cycle $\Delta \phi = \int dt A(t) + \int dt F_{\phi}(t)$.  The former term is a contribution to $\phic$, and may depend on, e.g., experimental variations in the current $I$.  In contrast, the latter term is $\phiq$, the phase diffusion due to spontaneous emission.  As the integral of white noise, $\phiq$ is a Gaussian random variable. This conclusion is not sensitive to the details of the model.  Rather, it is a consequence of our separation of the phase dynamics into $\phiq$, the part driven by spontaneous emission, and the part driven by everything else.   We do not estimate the amount of diffusion here, rather we leave this as a parameter, to study the relationship of phase diffusion to min-entropy generation, as in Figs.  \ref{fig:MinEntVsSig} and \ref{fig:MinEntVsSigBitDepth}.

From the phase invariance of Eqs. (\ref{eq:Pdot}) -- (\ref{eq:Ndot}), subsequent realizations of $\phiq$ are independent.  The phase invariance  is a possible weakness or point of  attack on the implementation.  If an adversary could introduce a coherent field at the laser frequency, they could bias the laser toward a chosen phase.  This attack appears difficult, however, as there is no optical connection to the outside world; all optical fibers terminate either on a photodetector or on an optical absorber.  In addition, in the implementation used here, an optical isolator incorporated into the laser package allows light to leave the laser, but not to enter it.  

}

\input{REUltrafast150113.bbl}

\end{document}

%% file: NewCommands.tex
\newcommand{\be}{\begin{equation}}
\newcommand{\ee}{\end{equation}}
\newcommand{\bea}{\begin{eqnarray}}
\newcommand{\eea}{\end{eqnarray}}

\newcommand{\nnequiv}{\nonumber \\ & \equiv &}

\newcommand{\nnp}{\nonumber \\ & & +}

\newcommand{\var}{{\rm var}}

%% file: REUltrafast150113.bbl
%